\documentclass[12pt]{revtex4}
\usepackage{textcomp}
\usepackage{setspace}
\usepackage{tipa}
\usepackage{verbatim} 
\usepackage{amsmath}
\usepackage{bbold}
\usepackage{subfigure}
\usepackage{amssymb}
\usepackage{accents}
\usepackage{amscd}
\usepackage{graphicx}

\input{Qcircuit}
\input epsf

\newcommand{\be}{\begin{equation}}
\newcommand{\nn}{\nonumber}
\newcommand{\ee}{\end{equation}}
\newcommand{\bea}{\begin{eqnarray}}
\newcommand{\eea}{\end{eqnarray}}

\newcommand{\bket}[1]{\left|#1\right\rangle\left\langle#1\right|}

\begin {document}
\title{Quantum walk with a four dimensional coin}

\author{Craig S. Hamilton$^1$, Aur\'{e}l G\'{a}bris$^1$, Igor Jex$^1$ and Stephen M. Barnett$^{1,2}$}
\address{$^1$Department of Physics
Faculty of Nuclear Sciences and Physical Engineering,
Czech Technical University in Prague,
B\v{r}ehov\'{a} 7,
Prague,
11519, Czech Republic \\ $^2$ SUPA, Department of Physics, University of Strathclyde, Glasgow G4 0NG, UK} 

%\eads{\mailto{igor.jex@fjfi.cvut.cz}}

%03.67.-a Quantum Information
%42.50.-p Quantum Optics
%42.50.Ex 	Optical implementations of quantum information processing and transfer 
%42.50.Tx 	Optical angular momentum and its quantum aspects

\begin{abstract}
We examine the physical implementation of a discrete time quantum walk with a four dimensional coin. Our quantum walker is a photon moving through a series of time delay loops, with time being our position space. Our quantum coin is the internal states of the photon: the polarization and two of the orbital angular momentum states. We demonstrate how to implement this physically and what components would be needed. We then illustrate some of the results that could be obtained from performing the experiment.
\end{abstract}

\maketitle

\section{Introduction}

Quantum walks \cite{Aharonov} have been studied in great detail now ever since it was shown that they are capable of performing algorithms faster than a  classical random walk \cite{Farhi98}. They have been used in search algorithms \cite{Shevi03, Gabris07} and recently a link to quantum computation was described \cite{Childs09,Kendon10}. For a review of the basic properties of quantum walks see \cite{Kempe03}.

There are two types of quantum walks: the continuous time quantum walk, specified by a Hamiltonian based on the adjacency  matrix of a graph which evolves simply by the Schr\"{o}dinger equation, and the discrete time quantum walk, which introduces an additional Hilbert space called a coin space which is ``flipped'' at each step and the particle movement depends upon the coin state. We focus on the discrete time case in this paper, briefly mentioning the main points next. 

The discrete time quantum walk evolves in steps, first by applying a coin operator, $\hat{C}$, to the coin space that ``flips'' the coin and then a step operator, $\hat{S}$, that moves the particle according to the coin state. One such step is given by,
\be
\ket{\phi_{t+1}}=\hat{S}\hat{C}\ket{\phi_t},
\ee
where $\ket{\phi_t}$ is the state of the particle at step `t', in position and coin space, and is of the form,
\be
\ket{\phi}_t =\sum_{n,j} \alpha_{n,j,t} \ket{n,j},
\ee
where $n$ is the particle position, $j$ is the coin state and the $\alpha_{n,j}$ are the amplitudes of the states. Coin dimensions have been studied with various operators to discover the dynamics of the state. One of the most frequently studied is a two dimensional coin where $\hat{C}$ is the Hadamard matrix,
\be
\hat{H}=\frac{1}{\sqrt{2}}\left [ \begin{array}{cc}
1 & 1\\
1& -1  \\
  \end{array} \right]\label{h2mat}.
\ee
The step operator $\hat{S}$ has the form
\be
\hat{S}=\sum_{n,j} \ket{n+e_j} \bra{n} \otimes  \bket{j}
\ee
where $e_j$ is the position shift due to the $j^{th}$ coin state. Typically unbiased walks are considered theoretically and experimentally, which are defined by $\sum_j e_j =0$.

Coins of more than two dimensions have been studied theoretically \cite{Brun03}. They can show complex dynamics depending upon the initial coin state and choice of coin operator. Localization has been shown to occur in quantum walks with multiple coins \cite{Inui05}. We can also have entanglement between two coins \cite{Venegas-Andraca05, Liu09}, which would be modelled by a four state system like ours. 

\begin{figure}[phtb]
\begin{center}
\scalebox{0.4}{\includegraphics[trim = 0mm 55mm 0mm 16mm, clip]{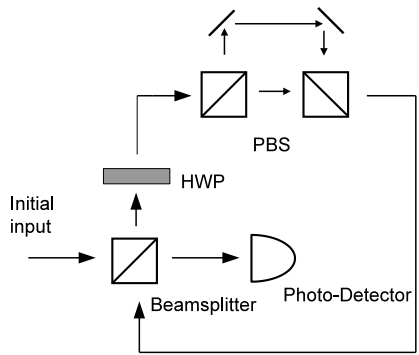}}
\end{center}
\caption{Experimental setup used by Schreiber {\it et.~al.} \cite{Jex10}.}\label{2dwalk}
\end{figure}

As well as the various theoretical studies there have also been experimental proposals and implementations of quantum walks. These include cold atoms in lattices \cite{Dur02, Eckert05}, ion traps \cite{Xue09} and recently the orbital angular momentum of photons \cite{Zhang10}. Previous work by Schreiber {\it et.~al.}~used an optical loop to realise a quantum walk with a two state coin by using the polarization of the photons \cite{Jex10}, and the setup is shown in figure~\ref{2dwalk}. We briefly explain this scheme here as we build upon it in this paper, replacing the coin and step operators used there with the ones we describe below for our extended coin space.

The initial state is produced by attenuating a coherent laser pulse down to the single photon level and rotating the polarization to $\ket{H}+i\ket{V}$, which will produce a symmetric distribution in position under the coin used, which is the Hadamard matrix, eq.~(\ref{h2mat}). This coin is realised as a half-wave plate (HWP) rotated at $22.5^\circ$ degrees. The step operator is a Mach-Zender interferometer with polarizing beam splitters (PBS) that send the polarizations along different routes that add different time delays to each polarization. At the end of the loop the photon has a probability to be out-coupled from the loop and be detected. This is a probabilistic implementation and the number of steps that the photon undergoes cannot be controlled.   
This experiment demonstrates impressively the typical features of quantum walks, such as ballistic spreading and interference.  

In this paper we propose an experimental scheme for a quantum walk with a four state coin by extending the coin space using the orbital angular momentum of the photon. This larger coin space will allow us to study more interesting dynamics and may open up applications for quantum information processing tasks to be performed. By modifying the initial state and/or the coin operator we can observe different dynamics of our quantum state, which may be important for performing quantum search algorithms.

In section~\ref{experimental_layout} we describe the experiment scheme of the walk, describing the physical implementation of the coin operator, step operator and initial state creation. In section~\ref{comp_results} we illustrate some of the results we could achieve from implementing this walk. Section~\ref{conc} is our conclusions.

\section{Experimental setup of 4-dimensional walk}\label{experimental_layout}

In this paper we propose an implementation of a quantum walk where the quantum coin has four states, which are composed of the polarization and two states of the photon's orbital angular momentum (OAM). The polarization states are the linear polarizations of light, horizontal and vertical (H,V). The OAM states of our coin space will be Laguerre-Gaussian modes \cite{Allen92}, which have an azimuthal angular dependence of $e^{il\phi}$ with $l =\pm 1$ (we label these $+,-$). We use these two states because they can be represented on a Poincar\'{e}-type  sphere on which we can visualise the state transformations in terms of rotations \cite{Padgett99}. Our four states are therefore $\ket{H,+}, \ket{H,-},\ket{V,+},\ket{V,-}$. We can relate these to the computational basis $\ket{00},\ket{01},\ket{10},\ket{11}$ and therefore associate the two attributes of the photon (polarization, OAM) to two qubits. 

Our physical implementation of the quantum walk consists of three parts: an initial state, a coin operator and the step operator. We describe each in turn below. The physical components of our experiment will be wave-plates and polarizing beam-splitters (PBS) for transforming the polarization and mode-converters \cite{Beijersbergen93} and dove prisms for transforming OAM. We combine these to realise the quantum coin operations we need. For example we can realise a controlled-NOT (CNOT) gate (where polarization is the control qubit and OAM the target) by a Mach-Zender interferometer using PBSs with a dove prism in one arm \cite{deOliveira05}.  The step operator displaces our photon throughout position space, which is time here.  All the steps are time delays and so we must change the midpoint of our walk with each step that occurs. In the following subsections we will describe the mathematical and physical forms of the individual coin and step operators. 

\subsection{Step Operator}

The step operator displaces each coin state by a different amount in our position space. To realise this physically we have to split the coin states spatially along different paths of an interferometer and then introduce different time delays to each path. In this section we comment on two possible realisations of the step operator, each with its own advantages and disadvantages. We first describe a component of both step operators; the Sagnac interferometer.

\subsubsection{Sagnac Interferometer}

\begin{figure}
\begin{center}
\scalebox{0.5}{\includegraphics[trim = 20mm 55mm 60mm 10mm, clip]{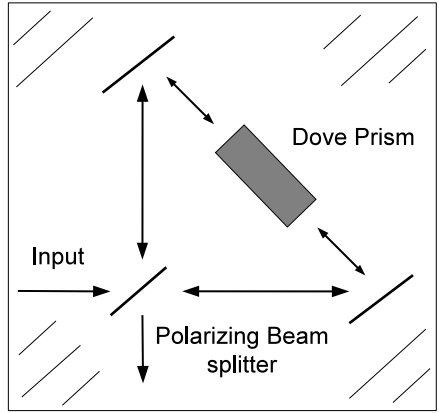}}%SIgimp.eps
\end{center}
\caption{Sagnac Interferometer. This device will be used later and represented as a hashed box in future diagrams.}\label{SI_fig}
\end{figure}

We wish to spatially separate the four coin states, and, while this is easy for the polarization qubit, it is difficult to do this coherently for the OAM state. To solve this problem we aim to transfer the quantum state encoded on the OAM sub-space to the polarization sub-space. To do this we utilise a Sagnac interferometer (SI) with a Dove prism (rotated at an angle $\theta$) in the path \cite{Nagali09b}, as shown in figure~\ref{SI_fig}. 
By using a Sagnac interferometer and single qubit polarization and OAM rotating elements we can achieve the transformation,
\be
\ket{H}  \otimes \left( \alpha\ket{+} +\beta\ket{-}\right )\rightarrow \alpha\ket{H,+}+\beta\ket{V,-},
\ee
with a similar transformation for the vertically polarized states.  The OAM states have become entangled with the polarization and can now be separated at a PBS. One important point about this transformation is that it will swap the amplitudes of two of the coin states ($\ket{H,-} \rightleftarrows \ket{V,-}$). The operation of the Sagnac interferometer and wave-plates/mode converters on the coin states is,
\be
\hat{U}_{\mbox{\scriptsize{SI}}}=\left [ \begin{array}{cccc}
1 & 0 & 0 & 0 \\
0& 0 & 0& 1 \\
0 & 0 & 1 & 0\\
0 & 1  & 0 & 0  \end{array} \right],
\ee
in the computational basis $[\ket{00} \ket{01} \ket{10} \ket{11}]^T$ or using the polarization/OAM basis states $[\ket{H+} \ket{H-} \ket{V+} \ket{V-}]^T$. Our evolution becomes,
\be
\ket{\phi_{t+1}}= \hat{U}_{\mbox{\scriptsize{SI}}}\, \hat{S}\,\hat{C}\,\ket{\phi_t},
\ee
which suggests we can simply absorb the operation, $\hat{U}_{\mbox{\scriptsize{SI}}}$, into our coin operator and also alter our initial state accordingly. Experimentally this makes the quantum walk more difficult to realise, as more qubit gates will be needed to correct the extra operator. We now describe how we can achieve this transformation enacted by the SI. 

The polarization states entering the Sagnac interferometer are rotated from H (V) to D (A) by a wave-plate where,
\bea
\ket{D} = \frac{1}{\sqrt{2}} \left ( \ket{H} +\ket{V} \right )\nn \\
\ket{A} = \frac{1}{\sqrt{2}} \left ( \ket{H} -\ket{V} \right )\nn.
\eea
The state enters the PBS and the H and V components traverse the interferometer in opposite directions. There is a Dove prism in the interferometer, rotated at an angle $\theta$, eliciting the changes in the states, 
\bea
\ket{H,\pm}&\rightarrow &e^{\mp2 il\theta }\ket{H,\mp} \nn\\
\ket{V,\pm}&\rightarrow &- e^{\pm2 il\theta }\ket{V,\mp}.\nn
\eea
Here $|l|=1$ so we set $\theta=\pi/8$. The state exiting the interferometer is then,
\bea
\alpha \ket{-} \left (a\ket{H}-ib\ket{V}\right) \nn\\
+\beta \ket{+} \left(a i \ket{H} - b\ket{V}\right)\nn
\eea
which, with $a=1, b=1$, can be written as,
\be
\alpha \ket{-} \ket{R} +i \beta\ket{+}\ket{L}.\label{stsi}
\ee
where R and L are the right and left circularly polarized light, 
\bea
\ket{R} = \frac{1}{\sqrt{2}} \left ( \ket{H} - i \ket{V} \right )\nn \\
\ket{L} = \frac{1}{\sqrt{2}} \left ( \ket{H} + i\ket{V} \right )\nn.
\eea
The `A' input polarization ($a=1,b=-1$) state yields a similar state to eq.~(\ref{stsi}) but R and L are interchanged. The polarization and OAM can then be rotated to the correct state with wave-plates and mode-covnerters. This transformation will feature in the next two sections where we describe the step operator in full.

\subsubsection{Step operator- version 1}

\begin{figure}[phtb]

\begin{center}

\scalebox{0.5}{\includegraphics[trim = 0mm 70mm 0mm 50mm, clip]{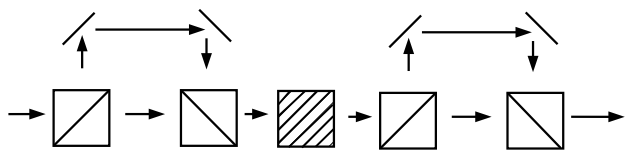}}
\end{center}
\caption{Physical implementation of step operator. There are four paths here. The dotted box represents a Sagnac interferometer and the boxes with diagonal lines representing polarizing beam splitters. The wave-plates and mode-converters that transform the photon polarization have been removed for clarity.}\label{step2}
\end{figure}

Our first realisation of the step operator is shown in figure~\ref{step2}.
The four coin states enter the step operator where one polarization has a time delay added to it by the first loop. We now entangle the OAM states with polarization, using the SI described in the previous section, and split the polarization again at another PBS. The advantages of this step operator are that is uses the minimum number of components which should reduce the losses and errors in the quantum walk. It has the disadvantage that the time delays added to the coin states are not completely independent for each coin state. 

It has the disadvantage that the time delays added to the coin states are not completely independent for each coin state. This is because the four time delays added are combinations of two loops and changing the time delays of these loops will change two of the four time delays.

\subsubsection{Step operator- version 2}

\begin{figure}[phtb]

\begin{center}

\scalebox{0.5}{\includegraphics[trim = 40mm 50mm  20mm 30mm, clip]{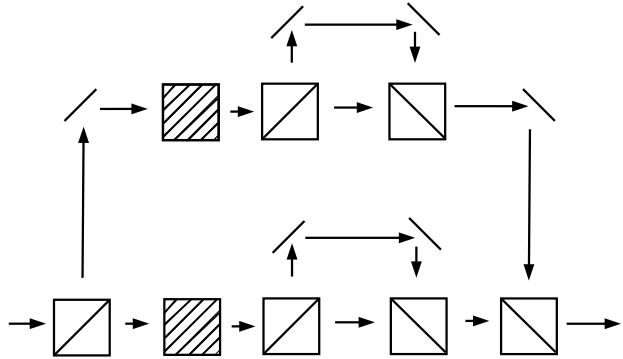}}
\end{center}
\caption{Physical implementation of step operator. There are four paths here. The dotted boxes represent Sagnac interferometers and the boxes with diagonal lines representing polarising beam splitters. The wave-plates that transform the photon polarization and OAM have been removed for clarity.}\label{step_long}
\end{figure}

We now present a different version of the step operator that has slightly different properties to the one described above. A layout of the step operator is shown in figure~\ref{step_long}.
It works on the same principle as the previous step operator but it has different paths for each polarization, firstly giving us four independent time delays and secondly we can have different wave plates in each part of the setup giving us different transformations on the coin states. This latter point gives us an advantage over the first version of the step operator, which we explain next. 

\subsubsection{Step operator- free space vs. optical fibres}

In the experiment performed by Schreiber {\it et.~al.} \cite{Jex10} they used optical fibres to introduce the time delay necessary for quantum walk as opposed to the free space propagation of the photon. We are not currently aware of any optical fibre that can preserve the OAM of the photons and therefore we either have to perform the experiment in free space or we can use another method to move in a zero OAM state, which we now describe. 

By using a device called a `q-plate' \cite{Marrucci06} we can move our photon from the $l=\pm1$ states into the $l=0$ state and thus send it through an optical fibre. The q-plate entangles photon polarization and OAM and the transformation for a q-plate with charge $q$ is,
\be
\ket{L}\ket{m}_{OAM} \rightleftarrows \ket{R}\ket{m+2q}_{OAM}.
\ee

We first say that we can only use this method in $2^{nd}$ version of the step operator, which will become clear soon. We introduce a q-plate with q=1/2 before and after our inner time delay loops in figure~\ref{step_long}. For this to work we require a specific state entering our q-plate,
\be
\alpha \ket{R, +} + \beta \ket{L,-} \xrightarrow{q-plate}  (\alpha \ket{L} + \beta \ket{R} ) \otimes \ket{0}_{OAM},\nn
\ee
because when this state passes through the q-plate it will be contained within the zero OAM subspace. We can now pass this state through an optical fibre. It can be shown that in the $1^{st}$ version of the step operator that we cannot obtain the correct state for both subspaces of the photon polarization, whereas in the 2nd version we have the flexibility to use different wave plates in each loop. After the state has passed through the optical fibres we use another q-plate to reverse the $1^{st}$ q-plates transformation, thus obtaining the initial state with the appropriate time delays. 

\subsection{Initial state creation}

We wish to create states of the form,
\be
\ket{\phi} = \alpha_1\ket{H,+}+ \alpha_2\ket{H,-}+ \alpha_3\ket{V,+}+ \alpha_4\ket{V,-},\label{ini_st}
\ee
from the initial state $\ket{H}\otimes\ket{0}$, a horizontally polarized photon with zero OAM. We do this by first converting the state $\ket{H,0}$ into one of the four basis states used, which can be accomplished by the use of wave-plates (to convert linearly polarized light into circular polarized light) and a q-plate with $q=1/2$. To transform this into a state of the form of eq.~\ref{ini_st} we require a general $U(4)$ operation although it may not need to be an entangling operation, making it easier to realise experimentally.

\subsection{Coin operator}

The coin operator is a U(4) matrix that transforms the coin states into one another. In this paper we will examine two coins: the Hadamard coin, $H_4$, and the Grover coin, $U_G$, although our scheme is flexible enough to implement any coin. The coin operators are given by the following matrices,
\be
\begin{array}{cccccc}
H_4 & = & \frac{1}{2}
\left [ \begin{array}{cccc}
1 & 1 & 1 & 1 \\
1 & -1 & 1 & -1 \\
1 & 1 & -1 & -1 \\
1& -1 & -1 & 1  \end{array} \right], &U_{G}&=&\frac{1}{2}
\left [ \begin{array}{cccc}
-1 & 1 & 1 & 1 \\
1 & -1 & 1 & 1 \\
1 & 1 & -1 & 1 \\
1& 1 & 1 & -1  \end{array} \right]  \end{array} \label{coinop}
\ee
Due to the swap operation in the step operator we have to modify our coin operators in the physical implementation of them. This involves multiplying the coin matrices by $\hat{U}^{-1}_{\mbox{\scriptsize{SI}}}$ and implementing that in the experiment. The gates we have at our disposal are single qubit gates (on both qubits) and the CNOT gate. We can use these to realise any U(4) operator \cite{Vidal04, Coffey08, Tucci}.

The Hadamard coin, $H_4$, is chosen because it has been studied in the two coin state and also because it has the form,
\be
H_4 = H^{\mbox{\scriptsize pol}}_2 \otimes H^{\mbox{\scriptsize{ OAM}}}_2
\ee
i.e. it will not entangle the two sub-spaces of the coin states and we can therefore study how initially entangled states evolve. The Hadamard coin can be simply realised as a half-wave plate followed by a $\pi-$mode converter, as shown in the quantum circuit diagram below,
\[
\Qcircuit @C=1.0em @R=0.7em  {
& \gate{H_2} & \qw &  && \multigate{2}{H_4} & \qw  \\
& & &=& && \\
& \gate{H_2} & \qw &                                                              &       & \ghost{H_4} & \qw
}
\]
where the top line represents the polarization qubit and the bottom line the OAM qubit. When we take into account our step operator, $\hat{U}_{\mbox{\scriptsize{SI}}}$,  the modified coin operator becomes,
\[
\Qcircuit @C=1.0em @R=0.7em  {
& \gate{H_2} &\ctrl{2}& \qw &  && \multigate{2}{H'_4} & \qw  \\
& & &&=& && \\
& \gate{H_2}&\targ & \qw &                                                              &       & \ghost{H'_4} & \qw
}
\]
where $H'_4=U^{-1}_{SI}H_4$.

The Grover coin is chosen as it has interesting evolutions depending on the input state \cite{Venegas-Andraca05}. It is possible to implement the modified Grover coin (taking into account $U_{\mbox{\scriptsize{SI}}}$) as four single qubit gates and one CNOT gate, as shown below in a circuit diagram,
\[
\Qcircuit @C=1.0em @R=.7em {
& \gate{U_1} & \ctrl{2} & \gate{U_2} & \qw &    &  &\multigate{2}{G'} & \qw\\
&                   &            &                   &        & = &  &                         & \\
& \gate{V_1} & \targ    & \gate{V_2} &  \qw      &    & &  \ghost{G'} & \qw
}
\]
The single qubit gates are,
\be
\begin{array}{cc}
U_1=\frac{e^{-i\pi/4}}{\sqrt{2}} \left [ \begin{array}{cc} 1 & 1 \\  -1 & 1  \end{array} \right] & U_2=\frac{e^{-i\pi/4}}{2} \left [ \begin{array}{cc} 1 & -1 \\ -i &  -i \end{array} \right] 
\end{array}\nn
\ee

\be
\begin{array}{cc}
V_1=\frac{e^{-i\pi/4}}{\sqrt{2}} \left [ \begin{array}{cc} i & 1 \\ -i & 1   \end{array} \right] & V_2=\frac{e^{-i\pi/4}}{\sqrt{2}} \left [ \begin{array}{cc} -i & -i \\ 1 &  -1\end{array} \right] 
\end{array}
\ee

In the next section we illustrate some of the evolutions that can be obtained from these coin operators.

\section{Computational results}\label{comp_results}

In this section we discuss the computational results of our quantum walk. We use the two coins mentioned above and our step sizes are $\pm1$ and $\pm 2$ and we evolve our walk for 12 steps as this is an experimentally realisable number. As all our steps are negative we re-scale our mid-point at each time step by moving the zero point back each time. In each figure we sum over all coin states and plot the probability distribution that the state is found at that position vs.~position. 

\begin{figure}[phtb]
\centerline {\epsfxsize=8.0cm \epsffile{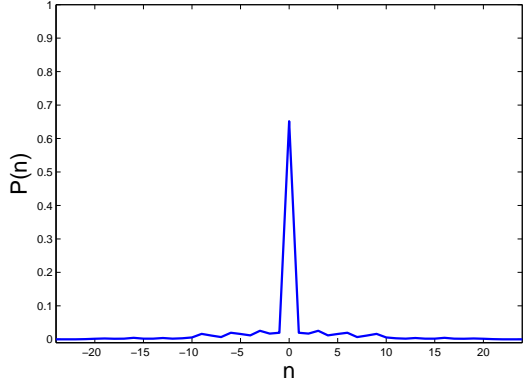}}
\caption{Quantum walk with 12 steps, using the Grover coin and the initial state $\ket{\phi}_{1}$. Plotted is the probability distribution $P(n)$ vs. position $n$.}\label{gcloc}
\end{figure}

\begin{figure}[phtb]
\centerline {\epsfxsize=8.0cm \epsffile{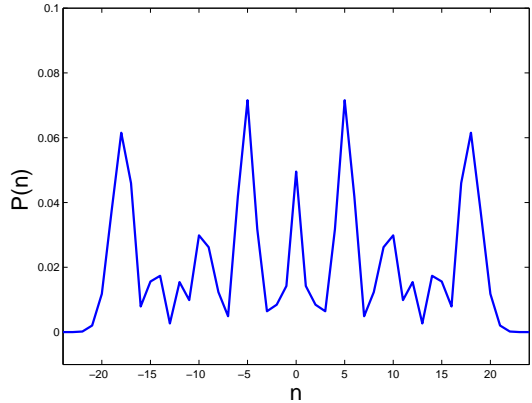}}
\caption{Quantum walk with 12 steps, using the Grover coin and the initial state $\ket{\phi}_{2}$. }\label{gc}
\end{figure}

In the first two figures, \ref{gcloc} and \ref{gc}, we use the Grover coin, $U_G$ from eq.~\ref{coinop}, with two different initial states,
\bea
\ket{\phi}_{1} &=& \frac{1}{2} \left [  \ket{H,+} + \ket{H,-}-  \ket{V,+} -\ket{V,-}  \right ]\\
\ket{\phi}_{2} &= &\frac{1}{2} \left [  \ket{H,+} + \ket{H,-} +  \ket{V,+} + \ket{V,-}  \right ],
\eea
and we can see very different behaviour in the evolution of the state. In figure~\ref{gcloc} we can see localization of the wave function about the midpoint and the marked difference compared to figure~\ref{gc}, which shows spreading over time. Note the difference in the scales of the vertical axes. This difference is due solely to the initial state used in the device and should be easy to demonstrate in an experimental setup. 

\begin{figure}[phtb]
\centerline {\epsfxsize=8.0cm \epsffile{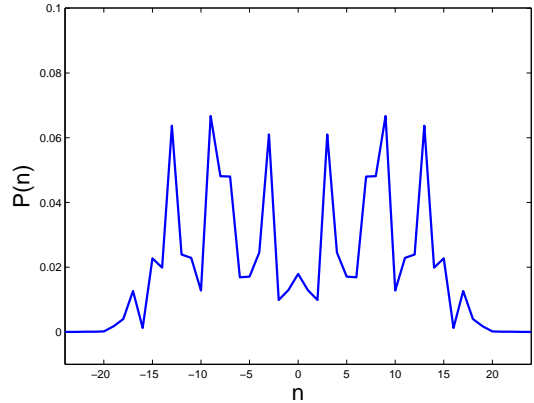}}
\caption{Quantum walk with 12 steps, using the Hadamard coin and the initial state $\ket{\phi}_{3}$.}
\label{hw}
\end{figure}

In figure~\ref{hw} we plot the evolution of the the Hadamard coin for 12 steps with the initial state,
\be
\ket{\phi}_{3}= \frac{1}{2} \left [  \ket{H,+} + \ket{H,-} + i\ket{V,+} + i\ket{V,-}  \right ]
\ee

\subsection{Two-dimensional walk}

\begin{figure}[phtb]
\scalebox{0.3}{\includegraphics[trim = 00mm 40mm  00mm 30mm, clip]{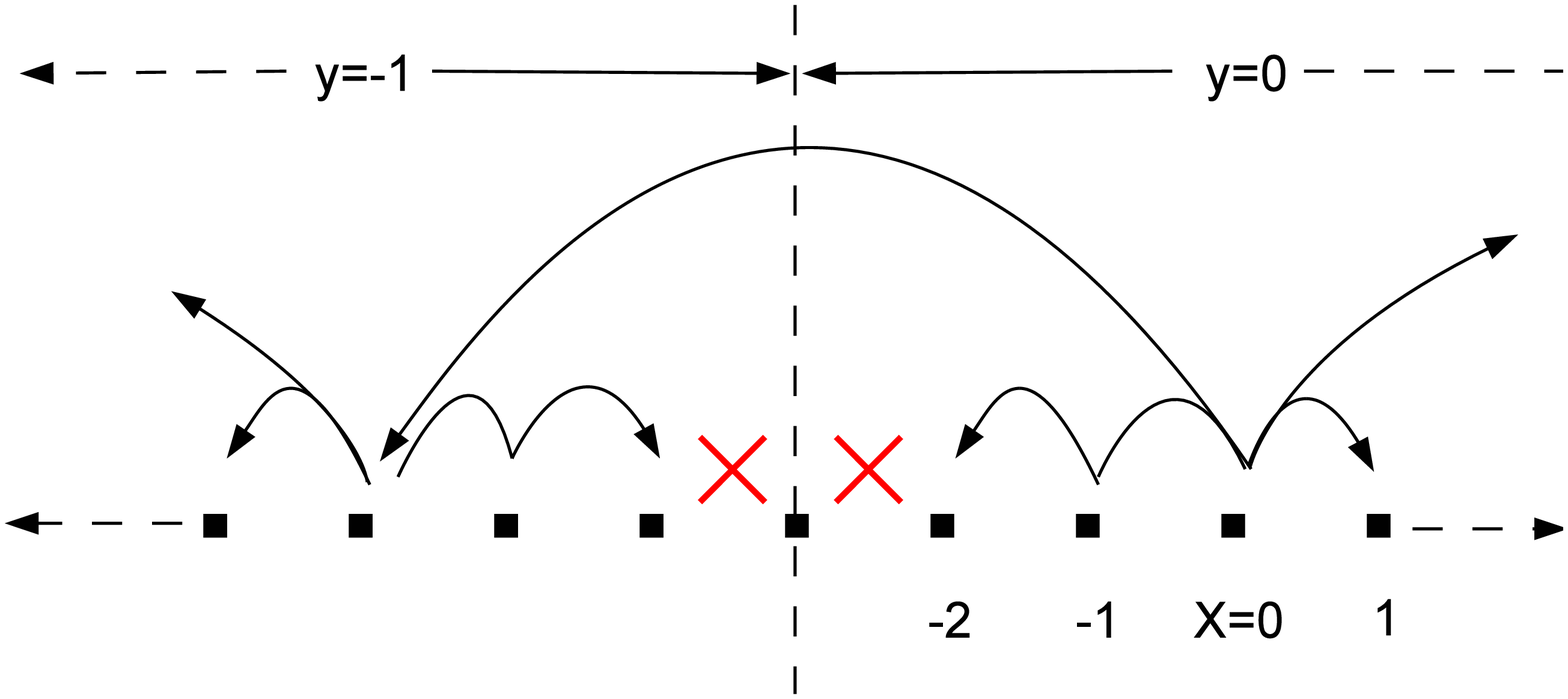}}
\caption{Two-dimensional quantum walk after two steps. There is a single point not occupied between the two sections of the walk and we can divide the x-axis into sections labelled $y=-2,-1,...,2$ (in this case). }\label{2dfig}
\end{figure}

In the next set of results we plot, the times delays are $\pm 1$ and $\pm N$, where $N$ is a large, odd integer. If we run this walk for a length of $(N-1)/2$ steps we can re-create the results of a two-dimensional walk by dividing the time axis up into segments, representing one of the dimensions (e.g. y), and within a segment is the other dimension (e.g x). It can be considered two-dimensional because there is zero overlap between segments. This is illustrated in figure~\ref{2dfig}.  
The results are shown below, where we chose $N=21$ and thus run each walk for 10 steps. If we assign each point on the x-axis a unique x,y co-ordinate then we can build up a two-dimensional plot of the quantum walk.

\begin{figure}
\centerline {\epsfxsize=8.0cm \epsffile{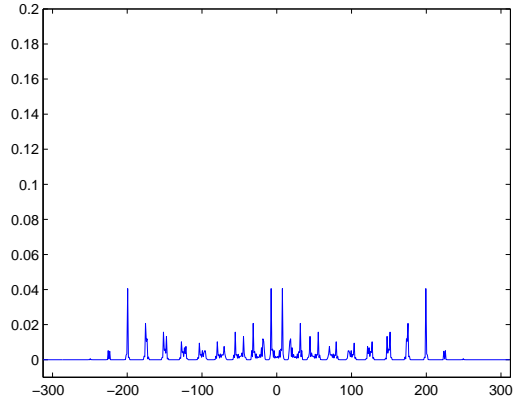}}
\caption{2D quantum walk with the Hadamard coin. Plotted is the probability distribution $P(n)$ vs. position $n$.}
\end{figure}

\begin{figure}
\centerline {\epsfxsize=8.0cm \epsffile{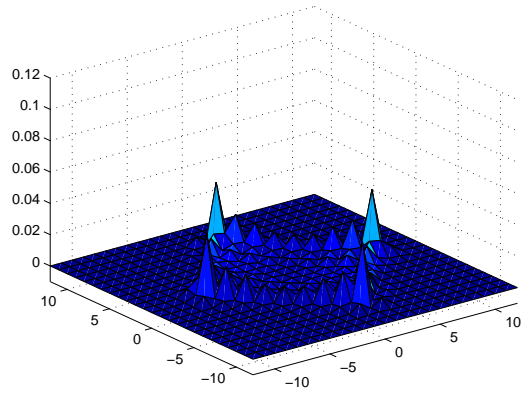}}
\caption{2D quantum walk with the Hadamard coin. By assigning each point on the time axis a point on an x,y grid we can build up a 2D picture of the walk. }
\end{figure}

\begin{figure}
\centerline {\epsfxsize=8.0cm \epsffile{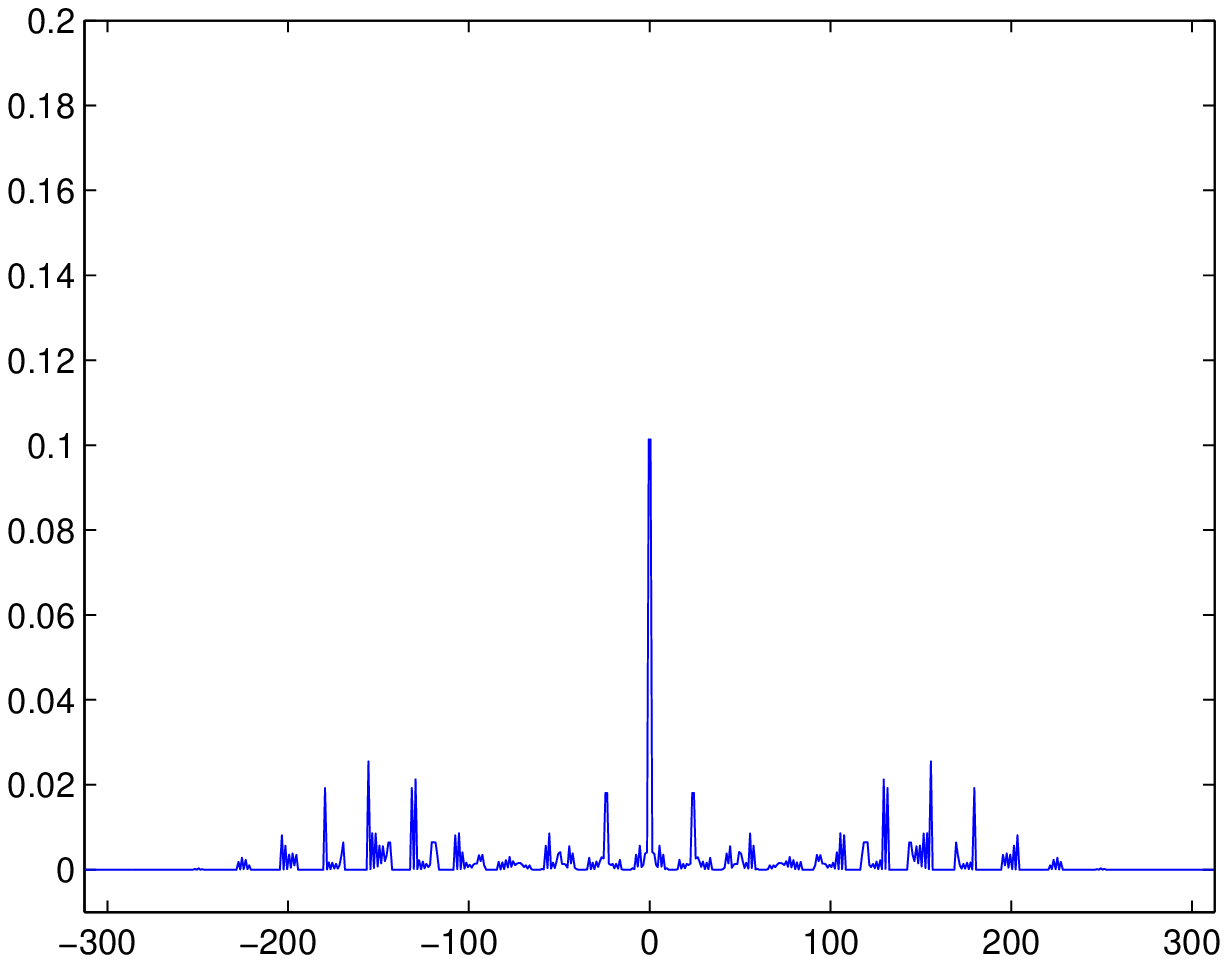}}
\caption{2D quantum walk with the Grover coin.}
\end{figure}
\begin{figure}
\centerline {\epsfxsize=8.0cm \epsffile{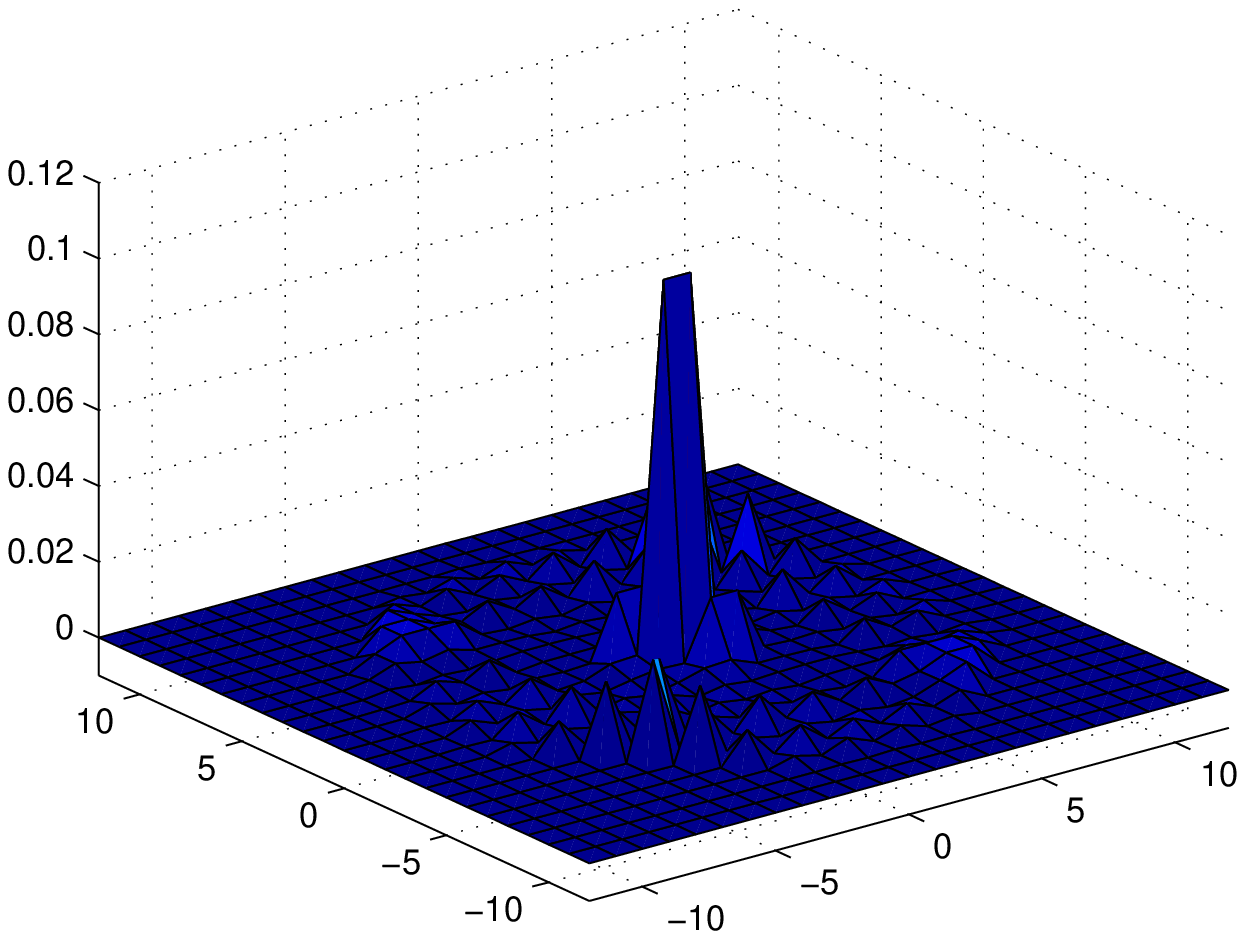}}
\caption{2D quantum walk with the Grover coin.}
\end{figure}

\section{Conclusions}\label{conc}

In this paper we have proposed an experimental realisation of quantum walk using a four dimensional coin. We described the physical implementation of the walk and pointed out some of the issues with this realisation. We have calculated some of the results that could be obtained from such a realisation for an experimentally realistic number of steps.

\section{Acknowledgements}
The author gratefully acknowledge helpful discussions with Prof.~C.~Silberhorn, Dr.~K.~Cassemiro and Mr.~A.~Schreiber. C.~S.~H.~, A.~G.~and I.~J.~acknowledge financial support from the Doppler Institute in Prague and from grants MSM6840770039
and MSMT LC06002 of the Czech Republic. S.~M.~B. thanks the Royal Society and the Wolfram Foundation for financial support.

\end{document}